# Topological geometric frustration in a cube-surface artificial spin ice

Zixiong Yuan[1,2,3], Wen-Cheng Yue[1,3,*], Peiyuan Huang[1,2,3], Yang-Yang Lyu[3], Sining Dong[1,3], Yi Dong[4,*], Huabing Wang[2,3], Peiheng Wu[2,3], and Yong-Lei Wang[1,2,3,*]

[1]*State Key Laboratory of Spintroincs Devices and Technologies, Nanjing University, Nanjing, China*

[2]*Purple Mountain Laboratories, Nanjing, China*

[3]*Research Institute of Superconductor Electronics, School of Electronic Science and Engineering, Nanjing University, Nanjing, China*

[4] *College of Metrology Measurement and Instrument, China Jiliang University, Hangzhou, China*

* Correspondence to: wenchengyue@nju.edu.cn, yingdong@cjlu.edu.cn, yongleiwang@nju.edu.cn

**Abstract**

**Artificial spin ices provide a controlled platform for investigating diverse physical phenomena, such as geometric frustration, magnetic monopoles, and phase transitions, via deliberate design. Here, we introduce a novel approach by developing artificial spin ice on the surfaces of a three-dimensional cube, which leads to emergent geometric frustration mediated by topologically protected domain walls, distinct from its flat counterparts. These domain walls connect vertices at the cube's corners that acting as intrinsic topological defects. Utilizing Monte Carlo simulations, we observe robust, topologically protected correlations among the intrinsic topological defects, regardless of their spatial separation. Our findings demonstrate that three-dimensional surfaces can unveil emergent properties absent in flat architectures.**

Geometric frustration is a compelling phenomenon that emerges when the geometric properties of a lattice prevent the simultaneous satisfaction of competing interactions [1–6]. Artificial spin ices (ASIs) have become prominent platforms for investigating geometric frustration in a controlled laboratory setting [5,7–12]. These systems are created using lithographically patterned arrays of nanoscale bar magnets. The interactions within ASIs can be precisely modeled using a dipolar model [7,13–15], facilitating a simplified and controlled exploration of the fundamental physics of magnetic frustration. So far, most investigations on ASIs have limited to two-dimensional (2D) flat structures. Quasi-three-dimensional (3D) ASI systems, featuring double layers of nanomagnets, have been investigated [16,17], they exhibit unique properties, such as extended degeneracy and emergent magnetic monopoles [16,17], distinct from their 2D analogues. Recent advances in sample fabrication techniques, including 3D nanofabrication [18–31] and self-assembly methods [32–35], have facilitated the creation of true 3D micro-/nano-structures, paving the way for the development of 3D ASIs and the exploration of their novel properties arising from the additional dimensional complexity. Consequently, interest in uncovering novel effects in 3D ASIs is on the rise [18–35].

In this letter, we propose a novel design of a 3D ASI consisting of a square ASI on the surface of a cube [Fig. 1(a)]. Our Monte Carlo (MC) simulations [36] reveal that while individual vertices are not subject to frustration, the entire structure exhibits emergent frustration due to the topology of the 3D surface structure. This leads to a significant increase in the system's degeneracy, coupled with the emergence of topologically protected domain walls, path-dependent phase frustration, and robust topological correlations. These findings underscore the complex interplay between topology and intriguing properties in 3D-surface ASIs.

The square ASI, a well-studied nanomagnet array, was originally designed to mimic the geometric frustration in pyrochlore spin ice [8]. This 2D planar structure features nanomagnets placed in a square lattice and exhibits a crystallized ground state with two-fold degeneracy [10]. In our study, we adapt the square ASI to a 3D context by placing it on the surface of a cube, as depicted in Fig. 1(a). On the cube's six faces, standard vertices of square ASIs are formed, each consisting of four moments and resulting in four types (I, II, III, and IV) of vertex configurations based on their energies [Fig. 1(c)]. Beyond these planar vertices, our cube-surface ASI incorporates two categories of 3D vertices located on the cube's edges and corners, respectively, as shown in Figs. 1(a) and 1(b). The 3D edge-vertex, also composed of four moments, mirrors the 16 vertex configurations (all structures can be found in Fig. S2 of Supplemental Material [37]) of the standard 2D square ASI. Its ground state, denoted as Type E-I, maintains a two-fold degenerate [Fig. 1(c)]. However, the excitation energy for Type II edge-vertices (Type E-II) is half that of the standard 2D Type II vertex (refer to Fig. 1c). Additionally, Type III edge-vertices ('three in, one out' or 'one in, three out') are divided into two groups based on their energies, labeled as Type E-III.a and Type E-III.b, with Type-III.a having lower energy than Type

E-III.b [see Fig. 1(c)]. The 3D corner-vertex consists of three moments, resulting in a six-fold degenerate ground state out of eight configurations, akin to that in traditional kagome (or honeycomb) ASIs.

To achieve the lowest energy ground state, ideally, all vertices in the system would align in their lowest energy state. However, our findings reveal that it is impossible for all vertices on the cube-surface square ASI to simultaneously achieve the lowest energy state. As illustrated by the thermal annealing results from our Monte Carlo simulations [Fig. 2(a)], a representative outcome selected from a total of 900 simulation results (derived from simulations of 9 distinct cube sizes, with 100 simulations conducted for each size) demonstrates that domain walls composed of Type F-II vertices (located on the faces) and Type E-II vertices (located on the edges) consistently manifest. Here, we do not discuss the impact of type III (marked as black dots) within the domain wall. Consider the spin configurations of three adjacent faces [Fig. 2(b)]: when all vertices on these faces are in their ground state Type I configurations, the 'spins' along the highlighted edge conflict [red in Fig. 2(b)], resulting in emergent frustration mediated by a domain wall. Given the closed, boundary-free nature of the cube's surface, the topology requires that the number of topologically protected domain walls across three adjacent faces must be odd, as illustrated in Figs. 2(a) and 2(c). In Fig. S3 of Supplemental Material [37], we show statistics on the number of domain walls connected to a corner vertex. These results reveal that the number of domain walls starting from a corner vertex must be odd.

Additionally, for the three surfaces converging at the same corner, at least one domain wall is required [Fig. 2c]. In a cube with six faces, a minimum of four domain walls are necessary for the ground state. Given that the excitation energy of Type E-II vertices (the energy gap between Type II and Type I vertices) on edges is lower than that of the standard Type II vertices on the faces, these four domain walls predominantly appear on the edges to achieve an ideal ground state, as depicted in Figure 2c. Monte Carlo simulations (refer to Video 1) demonstrate that these domain walls are located on two parallel faces of the cube (highlighted in Fig. S4 of Supplemental Material [37]), with each face containing domain walls on its parallel edges. A total of 288 different configurations meet these rules (details in the Supplementary Document [37] ,including references [38,39]), resulting in a ground state degeneracy of 288. This is significantly higher than the two-fold degeneracy observed in the conventional 2D planar square ASI, thus leading to topological frustration induced by the closed quasi-3D surface structure.

The corner-vertices in our 3D-surface square ASI act as topological defects, similar to the artificially introduced topological defects in 2D planar square ice [40]. Our MC simulations reveal that topological domain walls in the quasi-3D-surface square ASI invariably originate from one corner-vertex and terminate at another, as demonstrated in Fig. 3(a). This behavior contrasts with that in the 2D system, where, given the presence of boundaries, a domain wall can also start and end at the boundaries. Consequently, in 2D ices, the probability of topological defects not being interconnected by domain walls increases with the distance between defects, eventually reaching a probability of 1 [40]. This indicates that in the 2D structure, the correlation—defined as the probability that two topological defects are connected by domain walls—diminishes with increasing distance between the defects and approaches zero for sufficiently large separations. This suggests that distant topological defects in the planar 2D structure are essentially independent. In contrast, in the 3D cube-surface ASI, the topological defects at the corner-vertices remain correlated even regardless of their separation, as discussed below in Fig. 3b, ensuring a persistent and robust interconnectedness throughout the structure.

The separation of corner vertices in the 3D cube-surface ASI is determined by the cube's size. As there are no boundaries, each topological domain wall invariably connects two corner vertices (topological defects). We classify vertex pairs into three groups based on their separations within the cube: two corner-vertices on the same edge (connected by the domain walls indicated by purple dots in Fig. 3(a)), on face diagonals (connected by orange dots), and two on body diagonals (connected by green dots). To quantify the correlation between these topological defects, we calculate the probability that any two corner-vertices are connected by a topological domain wall, with results presented in Fig. 3(b). The data for each point in the figure is obtained by averaging the results of one hundred simulated annealing runs. We observe that the correlation between corner-vertices on the same edges is higher than those on face or body diagonals. Additionally, the correlation between corner-vertices on the same edges decreases with increasing cube size, as illustrated by the purple curve in Fig. 3(b). These are consistent with the reduction in correlation with defects' separation in 2D square ices [40]. Unlike in the 2D case, where distant topological defects become independent, in the 3D-surface ASI, the correlation between defects on the same edges diminishes gradually with increasing distance, reaching approximately 20% at large separations. Furthermore, correlations for corner-vertices on face diagonals and on body diagonals tend to increase with the cube's size. The average correlation among all corner-vertices remains nearly consistent (about 14.3%) and is unaffected by the cube's size, due to the minimum of four domain walls required by the topology of the 3D cube-surface ASI. These results indicate robust, topologically protected correlations among topological defects in the 3D surface ASI.

The correlation for the magnetic charge polarities of the corner vertices connected by a domain wall is complicated. As illustrated in Figs. 4(a), when two corner vertices are connected by a domain wall consisting of type-II square vertices, and there are no (or an even number of, as shown in Fig.4(c)) type III square vertices (which carry either a +2q or -2q charge), the polarities of the two corner vertices will be opposite. On the other hand, if the domain wall contains an odd number of type-III vertices, the charges of the two corner vertices will be the same, as illustrated in Figs. 4(b). In more complex scenarios, such as when a single corner-vertex connects to multiple other corner-vertices through several domain walls (as in Fig. 4(d)), the above rules may not hold. However, in general, the total net charge of all connected corner-vortices and type-III square-vortices must be zero.

In 2D planar square ices, the ground state typically exhibits a two-fold degenerated antiferromagnetic order [8]. The domain walls in this case, containing type II or III vertices, separate the two phases of the antiferromagnetic order. In our 3D cube-surface ASI, however, we observe an intriguing phenomenon of path-dependent phase frustration. As illustrated in Figs. 5(a) and 5(b), consider selecting an arbitrary vertex (highlighted by the red stars) and tracing a closed path loop on the surface originating from this vertex. When this loop encircles an odd number of topological defects at corner-vertices—such as the case with one corner-vertex in Fig. 5(a)—the path crosses domain walls an odd number of times (once in Fig. 5(a)). This results in an odd number of phase changes, leading to conflicted phases and resulting phase frustration when the path returns to the originating star vertex. However, the above phase frustration is path-dependent. As shown in Fig. 5(b), when a closed path loop encompasses an even number of topological defects, such as two corner-vertices, the phase at the starting vertex (indicated by a star in Fig. 5b) remains unchanged because the path cross domain walls an even number of times (twice in Fig. 5(b)). These observations highlight intriguing and peculiar phase changes depending on the specific path taken. Different closed paths originating from the same starting point can lead to different phases upon completion.

In this study, we employed Monte Carlo simulations to explore the emergent properties of square-ASI on the surface of a 3D cube, discovering a range of unique properties enabled by the addition of an extra dimension. We observed a significant increase in the ground state degeneracy, which fostered emergent topological frustration. The topological nature of the system induces the formation of topologically protected domain walls and ensures robust correlations between topological defects, regardless of their separation. It is possible to introduce additional artificial topological defects and/or corner-vertices in 3D surface ASI, as demonstrated in Fig. S5 of Supplemental Material [37]. These artificially introduced topological defects could modify the defects' correlation by interacting with intrinsic topological defects, potentially affecting the nucleation and movement of magnetic monopoles. The feasibility of 3D-surface ASI can be tested in future experiments through employing advanced 3D nanofabrication techniques [18–21,23,24,26–29] and self-assembly methods [32–35]. To date, the majority of 3D fabricated structures comprise interconnected networks. In these systems, the ground state of the connected square vortices predominantly adopts a type-II configuration, as demonstrated in several studies [23,31]. This may lead to distinct physical properties when compared to our proposed system, and we concur that further exploration of these interconnected networks presents a fascinating opportunity for future research. Additionally, while we recognize the challenges associated with fabricating purely dipolar-coupled nanomagnets in three dimensions, our proposed model could be more readily implemented using macroscopic artificial spin ice systems composed of bar magnets [41,42]. Furthermore, the topological frustration inherent in our 3D model may be more effectively realized through the use of 3D-printed mechanically frustrated systems [43,44]. Extending this approach to other ASI structures [45] such as kagome ASIs [46–48], direct-kagome ASIs [36], pinwheel

ASIs [49–51], and Shakti ASIs [5,6,52] on corresponding 3D polyhedron surfaces may reveal more intriguing physical properties. Artificial spin ice systems are renowned for their rich and diverse physical phenomena, which span a wide array of applications, including superconducting flux quanta [53–56], mechanical metamaterials [43,44,57], magnetic colloids [58,59], and qubits [60,61]. The complexity and richness of properties in the 3D surface ASIs could benefits diverse applications, including neuromorphic computing [62] and reconfigurable magnonics [63].

**Acknowledgments:** This work was supported by the National Natural Science Foundation of China (Grant Nos. 62288101, 62274086 and 12475042), the National Key R&D Program of China (Grant No. 2021YFA0718802), the Postdoctoral Fellowship Program of CPSF (Grant No. GZC20231108), the Jiangsu Funding Program for Excellent Postdoctoral Talent (Grant No. 2023ZB534), the China Postdoctoral Science Foundation under (Grant No. 2024M751370), and the Natural Science Foundation of Jiangsu Province (Grant No. BK20241223).

**Figures and Figure captions**

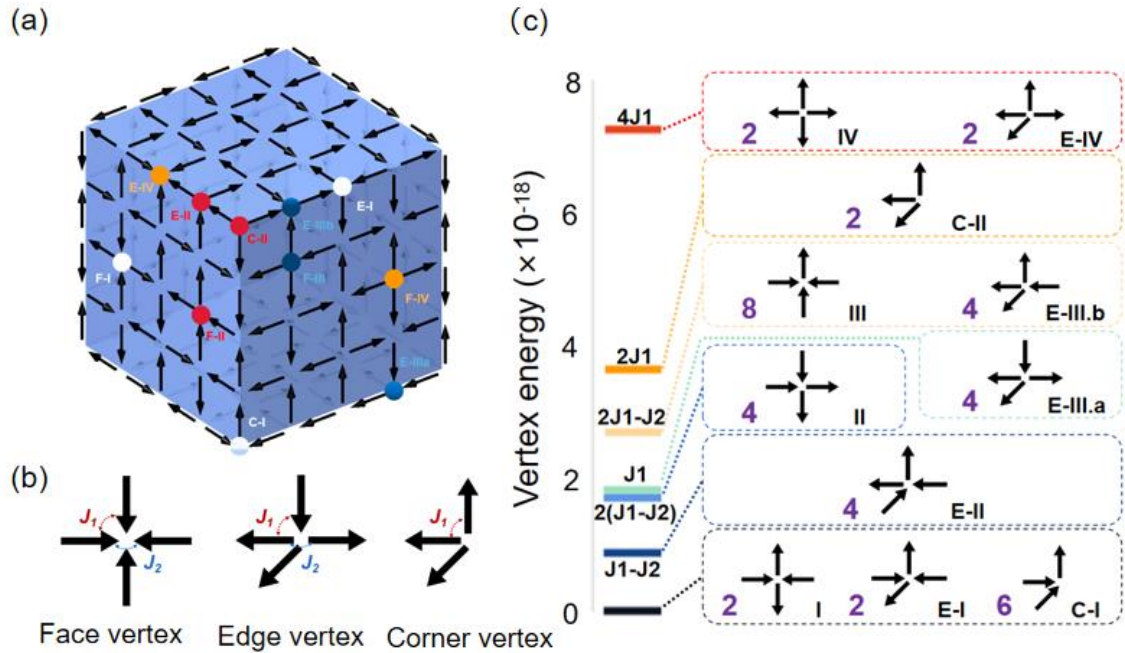

**Figure 1 | Three-dimensional cube-surface artificial spin ice.** (a) As shown in the figure, we define the size of a cubic structure with dimensions 4×4×4 as 4. Square ASIs on the surface of a cube, with artificial spins indicated by black arrows. Classified vertices (refer to Fig. c) based on energies are highlighted by dots with various colors. (b) Three vertex configurations with differing geometries. (c) Vertex classification based on energy levels. Purple numbers indicate vertex degeneracies; all degenerated configurations are detailed in the Figure S2 of Supplemental Material [37].

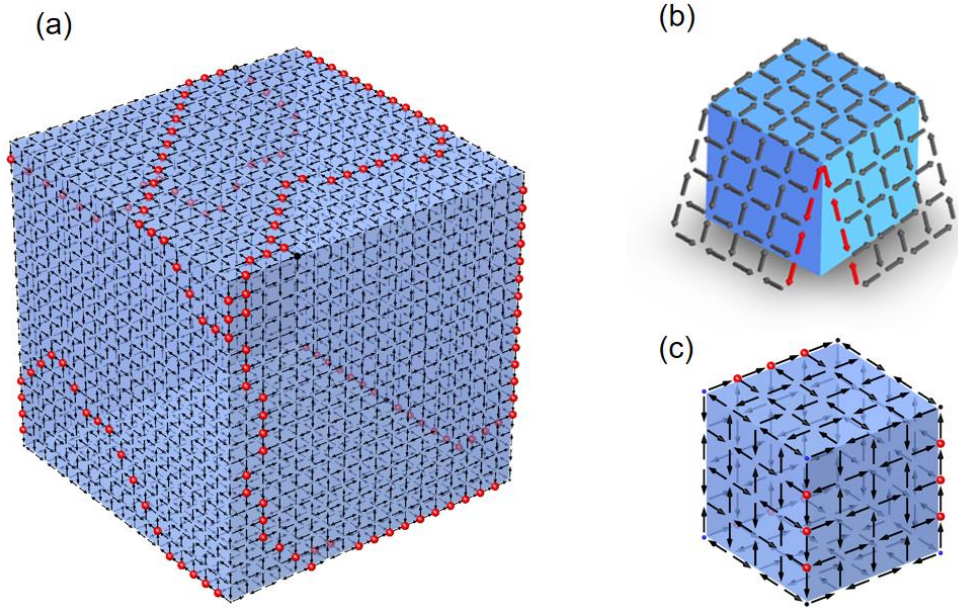

**Figure 2 | Emergent geometric frustration mediated by topologically protected domain walls.** (a) Spin configurations of thermal annealed model in MC simulations. Domain walls, composed of Type F-II vertices (on the faces) and Type E-II vertices (on the edges) are highlighted with red dots. (b) Ground state configuration of square spin ice across three unfolded adjacent faces. All vertices are in Type I ground state configurations. Highlight spins (in red) conflict at the junctions, leading to topological frustration. (c) A typical ground state configuration with a 288-fold degeneracy (refer to Fig. S3 of Supplemental Material [37]).

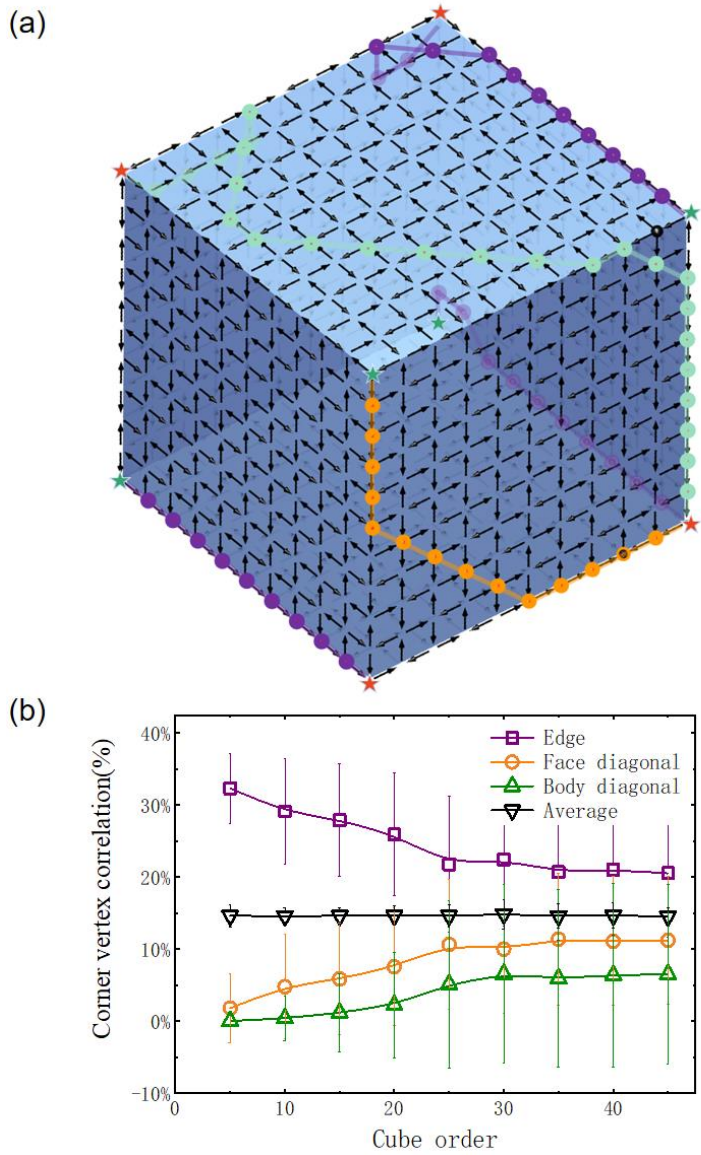

**Figure 3 | Topologically protected correlation of corner vertices.** (a) Correlation of corner-vertex pairs: Corner-vertex pairs on the same edges are connected by purple domain walls, pairs on face diagonals by orange domain walls, and pairs on body diagonals by green domain walls. The red star represents a positive charge, while a green star represents a negative charge. (b) Statistical analysis of corner-vertex correlation, defined by the probability that two corner-vertices are connected by domain walls.

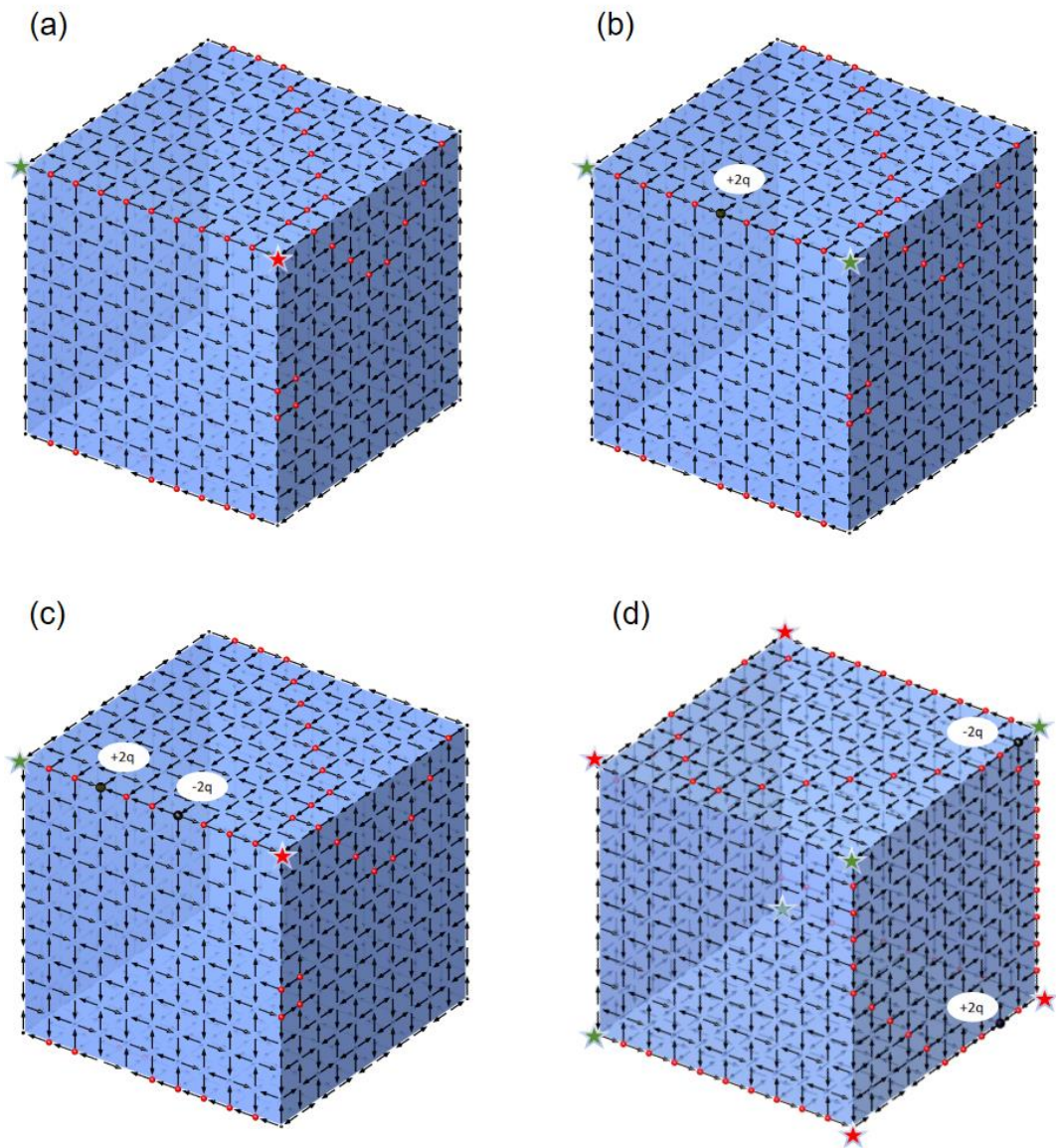

**Figure 4 | The correlation for the magnetic charge polarities of the corner vertices connected by a domain wall.** (a) When two corner vertices are connected by a domain wall consisting of type-II square vertices, and there are no type III square vertices, the polarities of the two corner vertices will be opposite. Here red stars represent positive charges, while green stars represent negative charges. (b) If the domain wall contains an odd number of type-III vertices, the charges of the two corner vertices will be the same. (c) Similar to the situation in Fig.a, if the domain wall contains an even number of type-III vertices, the charges of the two corner vertices will be opposite. (d) When a single corner-vertex connects to multiple other corner-vertices through several domain walls, the total net charge of all connected corner-vortices and type-III square-vortices must be zero.

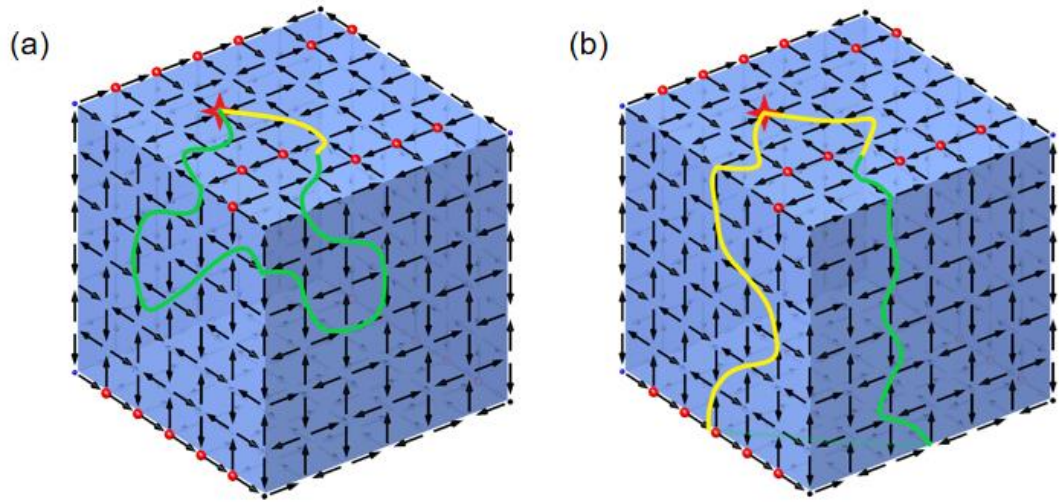

**Figure 5 | Path-dependent phase frustration.** (a) Path exhibiting phase frustration. The green and yellow paths represent two opposing phases of the Type I vertices they traverse. Vertices at domain walls are highlighted with red dots. The phase reverses each time it crosses a domain wall. (b) Path without phase frustration. The spin configurations in (a) and (b) are identical, illustrating the impact of path choice on phase frustration.

# Supplemental Materials

**Topological geometric frustration in a cube-surface artificial spin ice**

Zixiong Yuan[1,2,3], Wen-Cheng Yue[1,3,*], Peiyuan Huang[1,2,3], Yang-Yang Lyu[3], Sining Dong[1,3], Yi Dong[4,*], Huabing Wang[2,3,*], Peiheng Wu[2,3], and Yong-Lei Wang[1,2,3,*]

[1]*State Key Laboratory of Spintroincs Devices and Technologies, Nanjing University, Nanjing, China*

[2]*Purple Mountain Laboratories, Nanjing, China*

[3]*Research Institute of Superconductor Electronics, School of Electronic Science and Engineering, Nanjing University, Nanjing, China*

[4] *College of Metrology Measurement and Instrument, China Jiliang University, Hangzhou, China*

\* Correspondence to: wenchengyue@nju.edu.cn, yingdong@cjlu.edu.cn, yongleiwang@nju.edu.cn

**Monte Carlo simulations**

We performed our Monte Carlo (MC) simulations using a single spin flip algorithm. The interaction energies for these simulations were obtained from micromagnetic simulations with Mumax3 [38,39]. The nanomagnet size is 220 nm × 80 nm × 20 nm, and the lattice constant is 360 nm. We define the interaction between nearest neighbors with spins at a 90-degree angle as $J_1$, and the interaction between nearest neighbors with parallel spins as $J_2$. We chose the MC simulation parameters, such as the temperature range and the number of simulation steps (corresponding to thermal annealing time), to match the simulated ground state (Fig. S1) of a 2D flat square ASI with the experimental data reported in reference [36]. These same MC parameters were then used to simulate the 3D cube-surface square ASI. We only consider the interactions of the nearest neighbors. The thermal annealing protocol starts from a high-temperature $k_BT/J_2=10$. Each temperature update is set to 96% of the previous temperature value. Each temperature update includes 1000 simulation steps. In each step, calculations are performed for N/10 single-spin flips, where N represents the total number of spins in the system. The output of each step serves as the input for the next step.

**Ground state degeneracy**

The ground state configuration requires a minimal number of domain-wall vertices. For every set of three adjacent faces, at least one domain wall is needed due to the topological frustration arising from the 3D cube's structure (see Fig. 2(b)). A minimum of four domain walls is required for the whole cube. There are nine possible distributions for these four domain walls, as shown in Fig. S3(b).

Considering each domain has two configurations of Type E-II vertices (see Fig. S3(c)), there are $2^4=16$ possible cases for each domain wall distribution. Therefore, there are 9×16=144 possible configurations. Furthermore, since there are two phases of Type I orders, the total ground state degeneracy of the 3D cube-surface square ASI is 144×2=288.

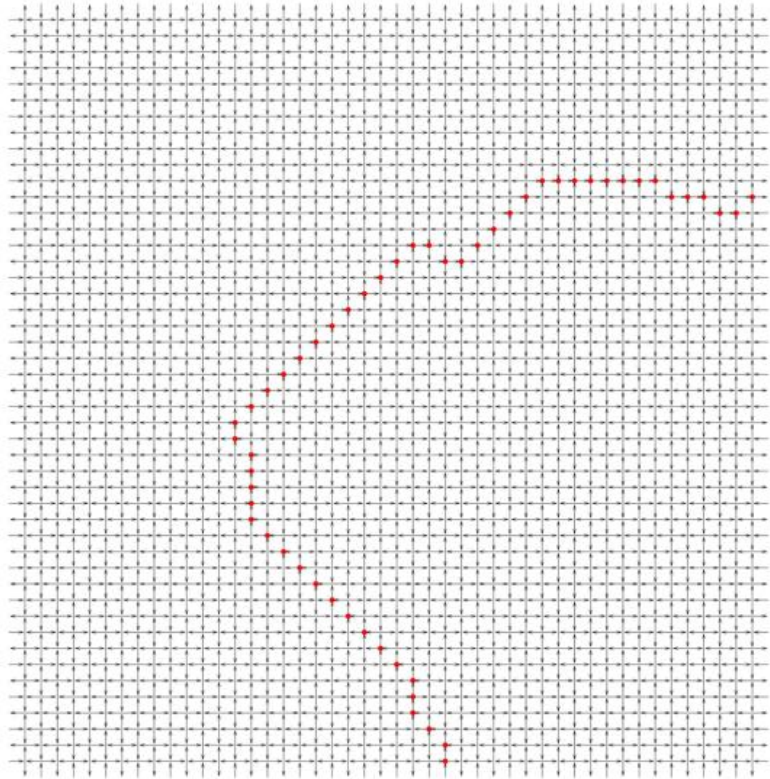

**Figure S1** | Monte Carlo simulation of the ground state configuration of a 2D square ASI resembles the experimental results in the supplemental document of reference [36].

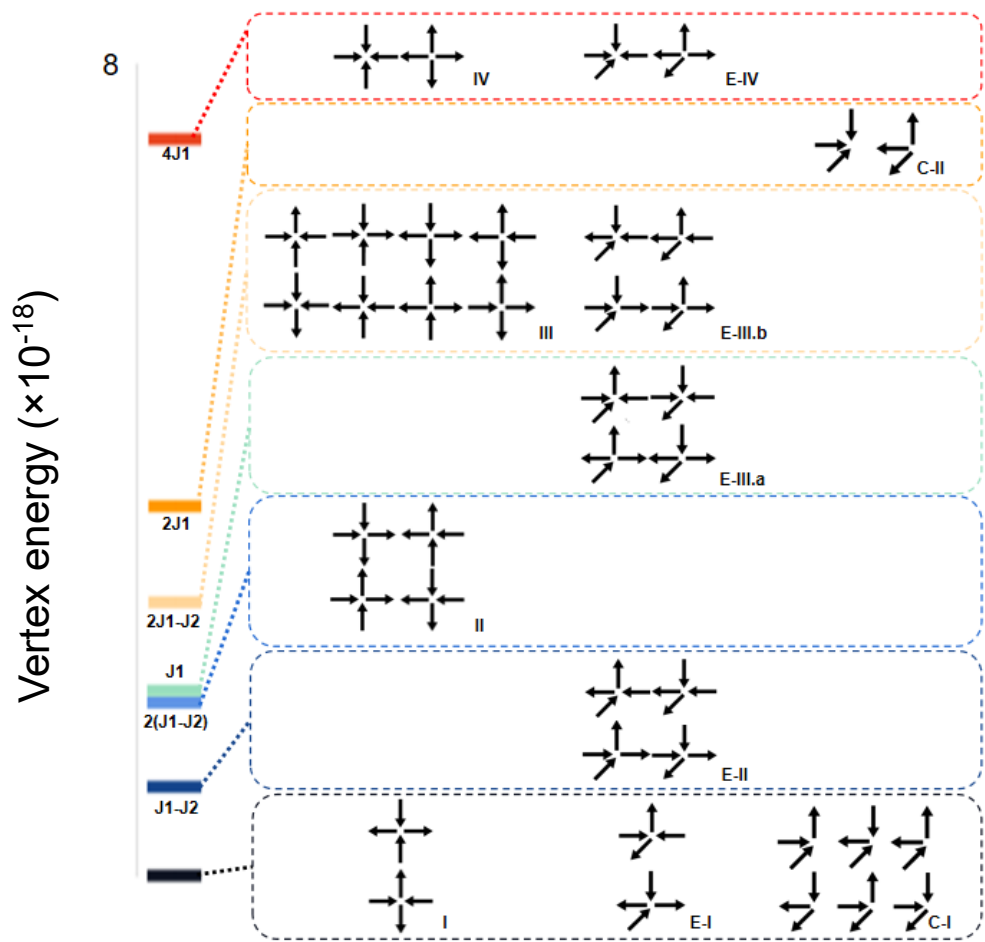

**Figure S2** | All possible vertex configurations of the 3D cube-surface square ASI.

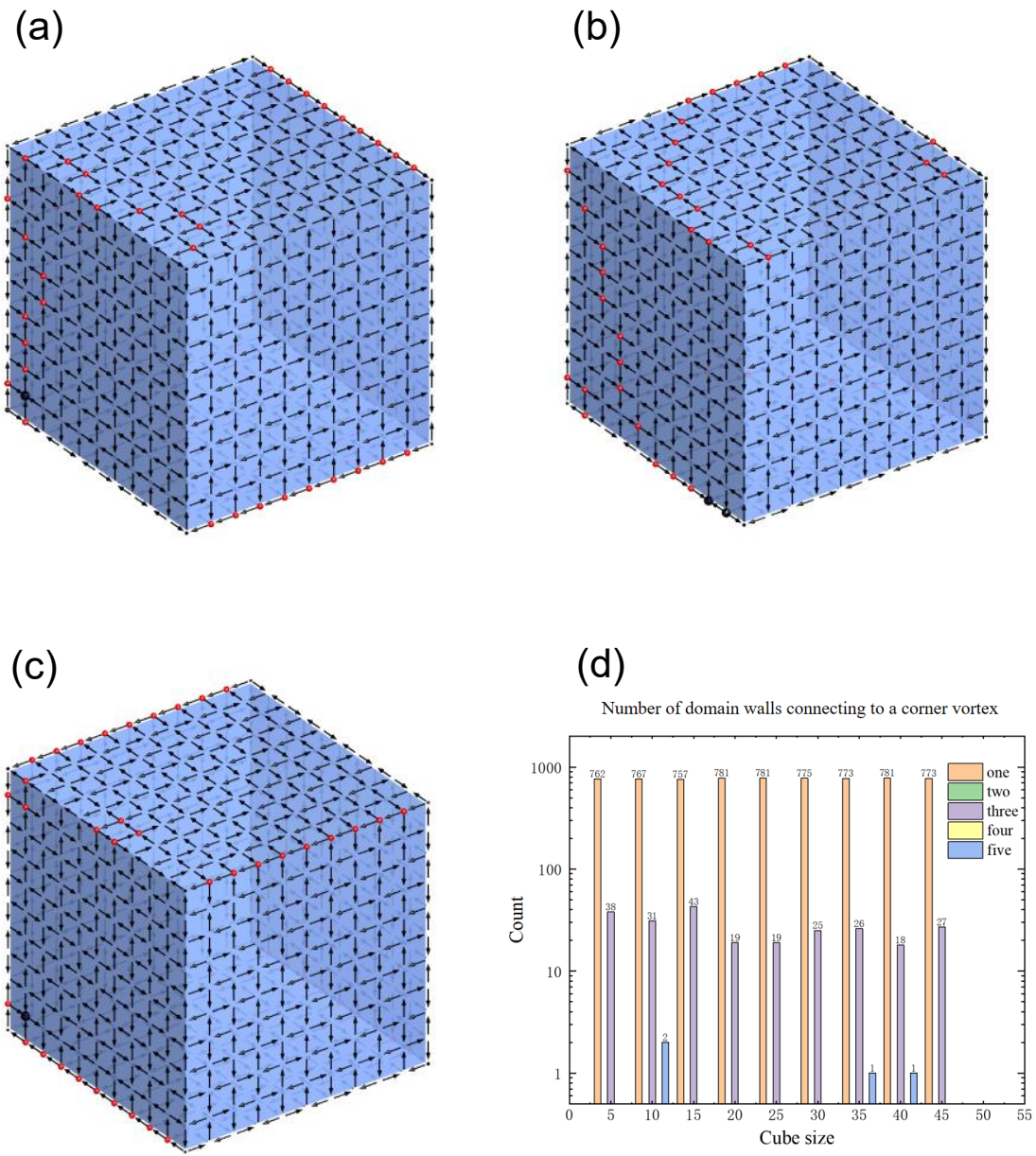

**Figure S3** | (a-c) Some other results selected from a total of 900 simulation results. (d) The statistical distribution of the number of domain walls connected to a corner vertex. Different colors represent number of domain walls connecting to the corner vortex.

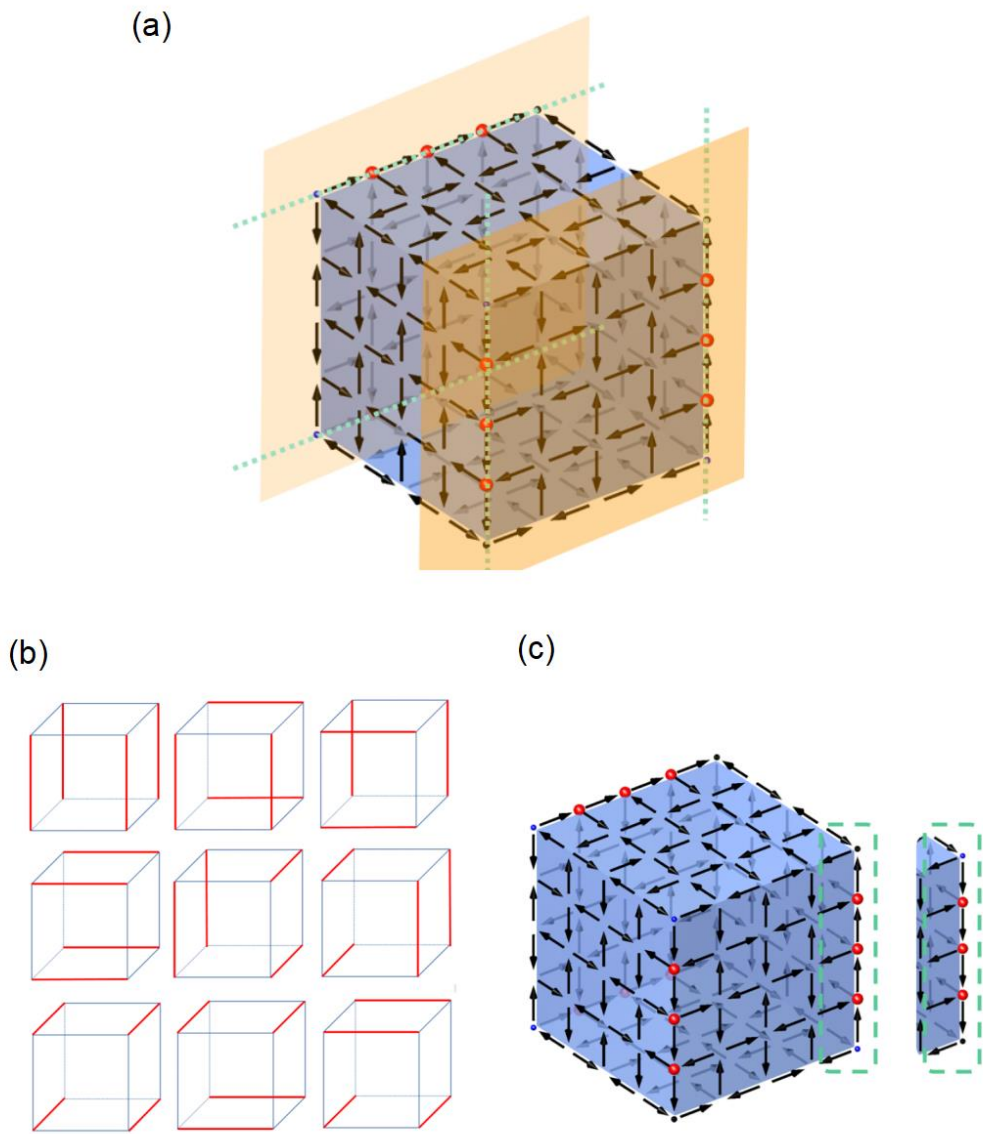

**Figure S4 |** (a) The four domain walls are situated on two parallel faces of the cube, each of which contains domain walls on its parallel edges. (b) Here are 9 possible domain walls distributions that satisfy the ground state requirements, each with a degeneracy of 32. (c) clarify that each domain wall can have two possible orientations.

(a)

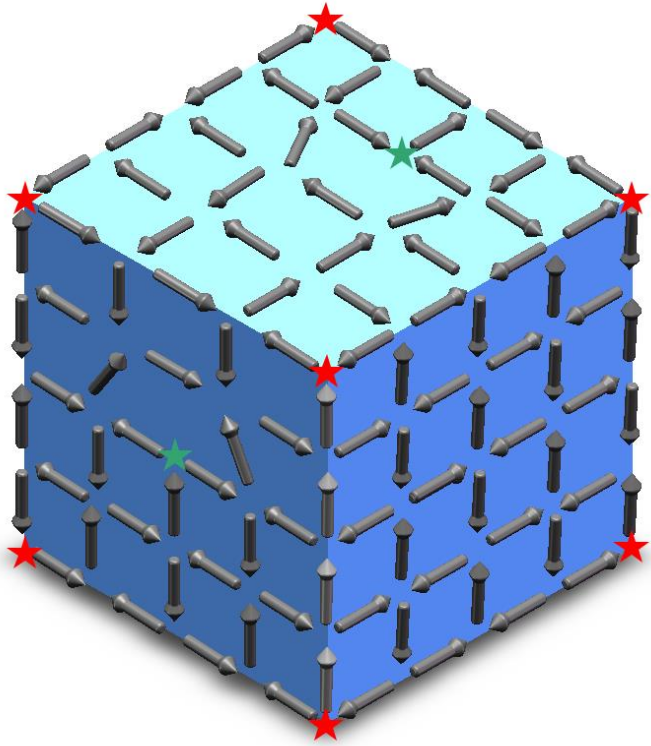

**Figure S5** | (a) Two artificially introduced topological defects (highlighted by green stars) on the 3D cube-surface square ASI.